\documentclass[twocolumn]{aastex63}

\usepackage{amsmath}

\received{September 8, 2020}
\revised{October 26, 2020}
\accepted{November 13, 2020}

\submitjournal{ApJL}

\shorttitle{}
\shortauthors{UNO \& MAEDA}

\graphicspath{{./}{figures/}}

\begin{document}

\title{Application of The Wind-Driven Model to A Sample of Tidal Disruption Events}

\correspondingauthor{KOHKI UNO}
\email{k.uno@kusastro.kyoto-u.ac.jp}

\author[0000-0002-6765-8988]{KOHKI UNO}
\affiliation{Department of Astronomy, Kyoto University, Kitashirakawa-Oiwake-cho, Sakyo-ku, Kyoto, 606-8502, Japan}

\author[0000-0003-2611-7269]{KEIICHI MAEDA}
\affiliation{Department of Astronomy, Kyoto University, Kitashirakawa-Oiwake-cho, Sakyo-ku, Kyoto, 606-8502, Japan}

\begin{abstract} 
An origin of the Optical/UV radiation from tidal disruption events (TDEs) has recently been discussed for different scenarios, but observational support is generally missing. In this Letter, we test applicability of the `Wind-Driven model' \citep{2020ApJ...897..156U} to a sample of UV/Optical TDEs. With the model, we aim to derive the physical properties of the Optical/UV TDEs, such as mass-loss rates and characteristic radii. The model assumes optically thick continuous outflows like stellar winds, and one key question is how the wind-launched radius is connected to physical processes in TDEs. As one possibility, through a comparison between the escape velocities estimated from their black-hole masses and the wind velocities estimated from observed line widths, we propose that the outflow is launched from the self-interaction radius ($R_{\rm SI}$) where the stellar debris stretched by the tidal force intersects; we show that the escape velocities at $R_{\rm SI}$ are roughly consistent with the wind velocities. By applying the model to a sample of Optical/UV TDE candidates, we find that explosive mass ejections ($\gtrsim 10 ~M_{\odot}{\rm yr^{-1}}$) from $R_{\rm SI}$ ($\sim 10^{14}{\rm ~cm}$) can explain the observed properties of TDEs around peak luminosity. We also apply the same framework to a peculiar transient, AT2018cow. The model suggests that AT2018cow is likely a TDE induced by an intermediate-mass black hole ($M_{\rm BH} \sim 10^{4}~M_{\odot}$).
\end{abstract}


\section{Introduction} \label{sec:1}

When a star approaches a supermassive black hole (SMBH) into its tidal radius, the tidal force of the SMBH destroys the star. These phenomena are known as Tidal Disruption Events (TDEs). In the TDE, roughly half of the stellar debris is bound and then accretes onto the SMBH, while the rest is unbound and escapes the system \citep{1988Natur.333..523R}. As a result, the system is believed to release a large amount of energy to power a bright transient for a relatively brief time ($\lesssim 1 {\rm ~yr}$).

In the classical picture, the accretion power is the main energy source of TDEs \citep{2002RvMA...15...27K}. The luminosity increases rapidly toward the peak, and then decreases with a power-law, described as $L \propto t^{-5/3}$ \citep{1989IAUS..136..543P}. The model predicts that the TDE has a peak in the radiation energy around the soft X-rays. Indeed, the first TDE candidate was discovered by ROSAT \citep{2002AJ....124.1308D}. However, in recent years, new generation surveys such as Pan-STARRS \citep{2002SPIE.4836..154K}, PTF \citep{2009PASP..121.1395L}, ASAS-SN \citep{2014ApJ...788...48S}, and ZTF \citep{2019PASP..131a8002B}, discover TDEs which are bright in the Optical/UV wavelengths. 

Typical features of the optical/UV TDEs include a high peak bolometric luminosity ($L_{\rm peak} \sim 10^{44}{\rm ~erg~s^{-1}}$), a high blackbody temperature at a few $\times 10^{4}{\rm ~K}$, a blue continuum component, and broad spectral lines of H, He, and N corresponding to the velocity of $\sim 10^{4}{\rm ~km~s^{-1}}$ \citep{2014ApJ...793...38A, 2020arXiv200101409V}. These features may depend on the physical properties of a disrupted star \citep[e.g.,][]{2012ApJ...757..134M}, a behavior of a stream stretched by the tidal force \citep[e.g.,][]{2009MNRAS.400.2070S}, radiative transfer effects \citep[e.g.,][]{2018ApJ...855...54R}, or other factors.

In recent years, it has been suspected that optically thick outflows of the stellar debris form the Optical/UV photosphere \citep{2009MNRAS.400.2070S, 2016MNRAS.461..948M}, and that the direct radiation from the disk is observable only in the late phase. As one origin of the outflows, some models have been proposed such as super-Eddington winds \citep{2009MNRAS.400.2070S, 2011MNRAS.410..359L} and the stream-collision outflow \citep{2016ApJ...830..125J, 2020MNRAS.492..686L}. However, the details still remain unclear.

In this study, by applying the `Wind-Driven model' \citep{2020ApJ...897..156U} to a sample of Optical/UV TDE candidates, we aim to understand the origin of the Optical/UV radiation. The model assumes optically thick continuous outflows characterized by the mass-loss rate ($\dot{M}$) and the wind velocity, which are analogous to stellar winds. We note that similar models have been indeed proposed for the Optical/UV TDEs \citep[e.g.,][]{2020ApJ...894....2P}, but so far the models have not been applied to a sample of observed TDEs \citep[but see][]{2020arXiv200901240M}. For example, in this Letter we show that the winds are likely to be launched from the self-interaction radii \citep[$R_{\rm SI}$, e.g.,][]{2015ApJ...806..164P, 2015ApJ...812L..39D}, through the comparison between escape velocities ($v_{\rm esc}$) and wind velocities ($v_{\rm wind}$). The radius is one of the characteristic radii of TDEs, in which the stellar debris stream stretched by the tidal force intersects. We further estimate the mass-loss rates of these TDEs at their peak luminosity and discuss the physical properties of the disrupted stars. 

The Letter is structured as follows. In Section\ref{fig:fig2}, we introduce observational properties of the sample of TDEs. In Section\ref{sec:3}, we apply the Wind-Driven model to these observed TDEs and compute their $R_{\rm SI}$. We also estimate the peak mass-loss rate ($\dot{M}_{\rm peak}$) for each TDEs. By applying the same framework to a peculiar transient, AT2018cow, we discuss its origin in Section\ref{sec:4}. The Letter is closed in Section\ref{sec:5} with conclusions.

\section{the properties of 21 TDEs}\label{sec:2}

We introduce some observational properties of the Optical/UV TDE candidates. We select the candidates discovered by surveys such as PTF, ASAS-SN, and ZTF. To obtain unique solutions using the Wind-Driven model, we need to select TDEs with sufficient information. The requirements are as follows; (1) the BH mass ($M_{\rm BH}$) has been estimated using the $M_{\rm BH}-\sigma$ relation or the $M_{\rm bulge}-M_{\rm BH}$ relation \citep{2013ApJ...764..184M,2013ARA&A..51..511K}, (2) the peak luminosity and temperature have been obtained, and (3) the spectral line widths have been derived around the peak luminosity. As the candidates that satisfy the above criteria, we select 21 Optical/UV TDEs. We summarize their observational properties in Table\ref{tab:TDEs}.

\begin{deluxetable*}{lcccccccccc}
\tablenum{1}
\tablecaption{Sample of 21 TDE candidates. \label{tab:TDEs}}
\tablewidth{0pt}
\tablehead{
\colhead{Object} & \colhead{$\log_{10}{M_{\rm BH}}$} & \colhead{$\log_{10}{L_{\rm peak}}$} & \colhead{$T_{\rm peak}$} & \colhead{spectral type} &
\colhead{} & \colhead{FWHM} & \colhead{} & \colhead{$L_{\rm edd}/L$} & \colhead{Ref.} & \colhead{$R_{\rm SI}$} \\
\cline{6-8}
 & [$M_{\odot}$] & [$\rm erg~s^{-1}$] & [$10^4{\rm ~K}$] & & \colhead{[$\rm km~s^{-1}$]} & line(\text{\AA}) & day & & & [$10^{14}{\rm ~cm}$] 
}
\startdata
      TDE2 & $7.00^{+0.30}_{-0.30}$ & $>43.6$ & $1.8$ & TDE-H & $8000$ & H$\alpha$ & - & 0.032 & a & 2.62\\
      PTF09ge & $6.31^{+0.39}_{-0.39}$  & $44.1$ & $2.2$ & TDE-He & $10100$ & \ion{He}{2}(4686) & $-1$ & 0.74 & b,c & 6.15\\
      PTF09axc & $5.68^{+0.48}_{-0.49}$ & $43.5$ & $1.2$ & TDE-H & $11900$ & H$\alpha$ & $+7$ & 0.50 & b,c & 3.94\\
      PTF09djl & $5.82^{+0.56}_{-0.58}$ & $43.9$ & $2.6$ & TDE-H & $6530$ & H$\alpha$ & $+2$ & 0.93 & b,c & 4.67\\
      PS1-10jh & $5.85^{+0.44}_{-0.44}$ & $44.5$ & $2.9$ & TDE-He & $9000$ & \ion{He}{2}(4686) & $<0$ & 3.1 & d,c,e & 4.82\\
      PS1-11af & $6.90^{+0.10}_{-0.12}$ & $43.9$ & $1.5$ & - & $10200$ & \ion{Mg}{2}(2680) & $+24$ & 0.082 & f,e & 3.14\\
      ASASSN-14ae & $5.42^{+0.46}_{-0.46}$ & $43.9$ & $2.2 $ & TDE-H & $17000$ & H$\alpha$ & $+3$ & 2.4 & g,c,e & 2.75\\
      ASSASN-14li & $6.23^{+0.39}_{-0.40}$ & $43.8$ & $3.5$ & TDE-Bowen & $3000$ & H$\alpha$ & $+10$ & 0.29 & h,c,e & 6.17\\
      ASASSN-15lh & $8.88^{+0.60}_{-0.60}$ & $45.3$ & $2.1$ & - & $7300$ & $4200$ & - & 0.020 & i & 0.887\\
      ASASSN-15oi & $6.40^{+0.54}_{-0.55}$ & $43.1$ & $2.0$ & TDE-He & $20000$ & \ion{He}{2}(4686) & $+7$ & 0.57 & j,e & 5.97\\
      iPTF15af & $6.88^{+0.38}_{-0.38}$ & $44.2$ & $4.9$ & TDE-Bowen & $11000$ & \ion{He}{2}(4686) & $+7$ & 0.15 & k,c & 3.28\\
      iPTF16axa & $6.34^{+0.42}_{-0.42}$ & $44.0$ & $3.0$ & TDE-Bowen & $8800$ & H$\alpha$ & $+6$ & 0.38 & l,c,e & 6.11\\
      iPTF16fnl & $5.50^{+0.42}_{-0.42}$ & $43.0$ & $2.1$ & TDE-Bowen & $10000$ & H$\alpha$ & $0$ & 0.24 & m,c,e & 3.09\\
      AT2017eqx & $6.77^{+0.17}_{-0.18}$ & $44.6$ & $2.1$ & TDE-Bowen & $19000$ & H$\alpha$ & $+11$ & 0.43 & n,e & 3.94\\
      PS18hk & $6.90^{+0.29}_{-0.30}$ & $43.9$ & $1.5$ & TDE-H  & $11500$ & H$\alpha$ & $+6$ & 0.085 & o,e & 3.16\\
      ASASSN-18jd & $7.60^{+0.40}_{-0.40}$ & $44.7$ & $2.9$ & TDE-Bowen & $3250$ & H$\alpha$ & average & 0.087 & p & 0.782\\
      ASASSN-18pg & $6.99^{+0.23}_{-0.23}$ & $44.4$ & $3.1$ & TDE-Bowen & $15000$ & H$\alpha$ & - & 0.17 & q,e & 2.67\\
      AT2018hyz & $6.09^{+0.30}_{-0.30}$ & $44.3$ & $2.2$ & TDE-H & $17000$ & H$\alpha$ & peak & 1.2 & r,e & 5.89 \\
      ASASSN-19bt & $6.78^{+0.26}_{-0.26}$ & $44.1$ & $1.9$ & TDE-H & $27000$ & H$\alpha$ & $+8$ & 0.15 & s,e & 3.89\\
      ASASSN-19dj & $7.10^{+0.22}_{-0.22}$ & $44.8$ & $4.5$ & TDE-Bowen & $17000$ & H$\alpha$ & - & 0.38 & t,e & 2.17 \\ 
      AT2019qiz & $5.75^{+0.45}_{-0.45}$ & $43.7$ & $1.9$ & TDE-Bowen & $15000$ & H$\alpha$ & - & 0.64 & u,e & 4.30 \\
\enddata
\tablecomments{We show the observed properties ($M_{\rm BH}$, $L_{\rm peak}$, $T_{\rm peak}$, and FWHM) for a sample of 21 TDEs studied in this Letter. The spectral type refers to \citet{2020arXiv200101409V}. In the 8th column, we list the date when their FWHM was observed since their peak luminosity. We calculate $R_{\rm SI}$ assuming $R_{*} = R_{\odot}$, $M_{*} = M_{\odot}$, and $\beta = 1$ (see the main text in Section\ref{sec:3.1}).\\
(References) a: \citet{2011ApJ...741...73V}, b: \citet{2014ApJ...793...38A}, c: \citet{2017MNRAS.471.1694W}, d: \citet{2012Natur.485..217G}, e: \citet{2020ApJ...894L..10H}, f: \citet{2014ApJ...780...44C}, g: \citet{2014MNRAS.445.3263H}, h: \citet{2016MNRAS.455.2918H}, i: \citet{2016NatAs...1E...2L}, j: \citet{2016MNRAS.463.3813H}, k: \citet{2019ApJ...873...92B}, l: \citet{2017ApJ...842...29H}, m: \citet{2017ApJ...844...46B}, n: \citet{2019MNRAS.488.1878N}, o: \citet{2019ApJ...880..120H}, p: \citet{2020MNRAS.494.2538N}, q: \citet{2020ApJ...898..161H}, r: \citet{2020MNRAS.498.4119S}, s: \citet{2019ApJ...883..111H}, t: \citet{2020MNRAS.tmp.2967H}, u: \citet{2020MNRAS.499..482N}}
\end{deluxetable*}

In Table\ref{tab:TDEs}, we present the classification based on the observed spectra presented in \citet{2020arXiv200101409V}; TDE-H, TDE-Bowen, and TDE-He. We also show the Full-Width Half Maximum (FWHM) of a selected line. We use the FWHM as their $v_{\rm wind}$. We use H$\alpha$ for objects where H$\alpha$ is observed, i.e., TDE-H and TDE-Bowen, and \ion{He}{2} instead for those where H$\alpha$ is not observed, i.e., TDE-He. PS1-11af has featureless spectral lines and we cannot identify the spectral lines. We do not also identify the classification of ASASSN-15lh. As $v_{\rm wind}$ of PS1-11af, we use the spectral line width around 2680\text{\AA}, which is presumed to be \ion{Mg}{2} \citep{2014ApJ...780...44C}. We use the width of the main feature at 4200\text{\AA} for $v_{\rm wind}$ of ASASSN-15lh \citep{2016NatAs...1E...2L}.

\section{Wind-Driven Model for a Sample of TDEs}\label{sec:3}

In some models, it is believed that optically thick flares or outflows originated in stellar debris form in the Optical/UV photosphere and emission and/or absorption in their spectral lines \citep{2009MNRAS.400.2070S, 2016MNRAS.461..948M,2020MNRAS.492..686L,2020ApJ...894....2P}. Here, we consider the Wind-Driven model by \citet{2020ApJ...897..156U} to test whether continuous outflows can explain the properties of the Optical/UV TDEs.

We apply the Wind-Driven model to the sample of the 21 TDE candidates. By applying the model, we can estimate some physical properties (the mass-loss rates and some physical scales) from observational properties (luminosity, temperature, and wind velocities). The model defines the innermost radius ($R_{\rm eq}$) where the wind is launched. In the original formalism, it is assumed that equipartition is realized between the internal energy (dominated by radiation) and the kinetic energy. However in this Letter, we follow an inversed approach; we first assume $R_{\rm eq} = R_{\rm SI}$, and then test whether the equipartition is indeed realized there. This way, we will show below that this assumption is supported by TDE observations.

\subsection{The Assumption: $R_{\rm eq} = R_{\rm SI}$}\label{sec:3.1}

$R_{\rm SI}$ depends on physical properties of the BH and the disrupted star. It is described by \citet{2015ApJ...812L..39D} and \citet{2017MNRAS.471.1694W} as follows:
\begin{align}
    R_{\mathrm{SI}}=\frac{R_{\mathrm{t}}(1+e)}{\beta(1-e \cos (\delta \omega / 2))},
\end{align}
where $R_{\rm t}$ is the tidal radius given as $R_{\rm t} \approx R_{*} (M_{\rm BH}/M_{\rm *})^{1/3}$, where $R_{\rm *}$ and $M_{*}$ are the radius and mass of the disrupted star. The impact parameter is given as $\beta = R_{\rm t}/R_{\rm p}$, where $R_{\rm p}$ is the pericenter distance. $e$ is the orbital eccentricity. $\delta \omega$ is given by \citet{2017MNRAS.471.1694W} as follows:
\begin{align}
    \delta \omega=A_{\mathrm{S}}-2 A_{\mathrm{J}} \cos (i),
\end{align}
where $i$ is the inclination. $A_{\rm S}$ and $A_{\rm J}$ are given by \citet{2010PhRvD..81f2002M} as follows:
\begin{align}
A_{\mathrm{S}}&=\frac{6 \pi}{c^{2}} \frac{G M_{\mathrm{BH}}}{R_{\mathrm{p}}(1+e)}, \quad {\rm and} \\
A_{\mathrm{J}}&=\frac{4 \pi a_{\mathrm{BH}}}{c^{3}}\left(\frac{G M_{\mathrm{BH}}}{R_{\mathrm{p}}(1+e)}\right)^{3 / 2},
\end{align}
where $c$, $G$, and $a_{\rm BH}$ are the light speed, the Newtonian constant of gravitation, and the BH spin, respectively.

For TDEs, the condition, $\beta \gtrsim 1$, needs to be satisfied. In this study, we assume $\beta = 1$. This assumption is appropriate in comparing the model with the observations, since $\beta = 1$ means a large collision cross-section, i.e., a high event rate. In addition, a low $\beta$, i.e., a low angular momentum, is likely preferred to produce low energy radiation such as the Optical/UV wavelengths \citep[e.g., see][]{2015ApJ...812L..39D}. We also assume that the radii and masses of the disrupted stars are $R_{*} = R_{\odot}$ and $M_{*} = M_{\odot}$. This assumption would be acceptable, as the main-sequence stars like the sun are most likely to be destroyed since their abundance. Under these assumptions, we compute $R_{\rm SI}$ as shown in Table\ref{tab:TDEs}.

\begin{figure}
\epsscale{1.17}
\plotone{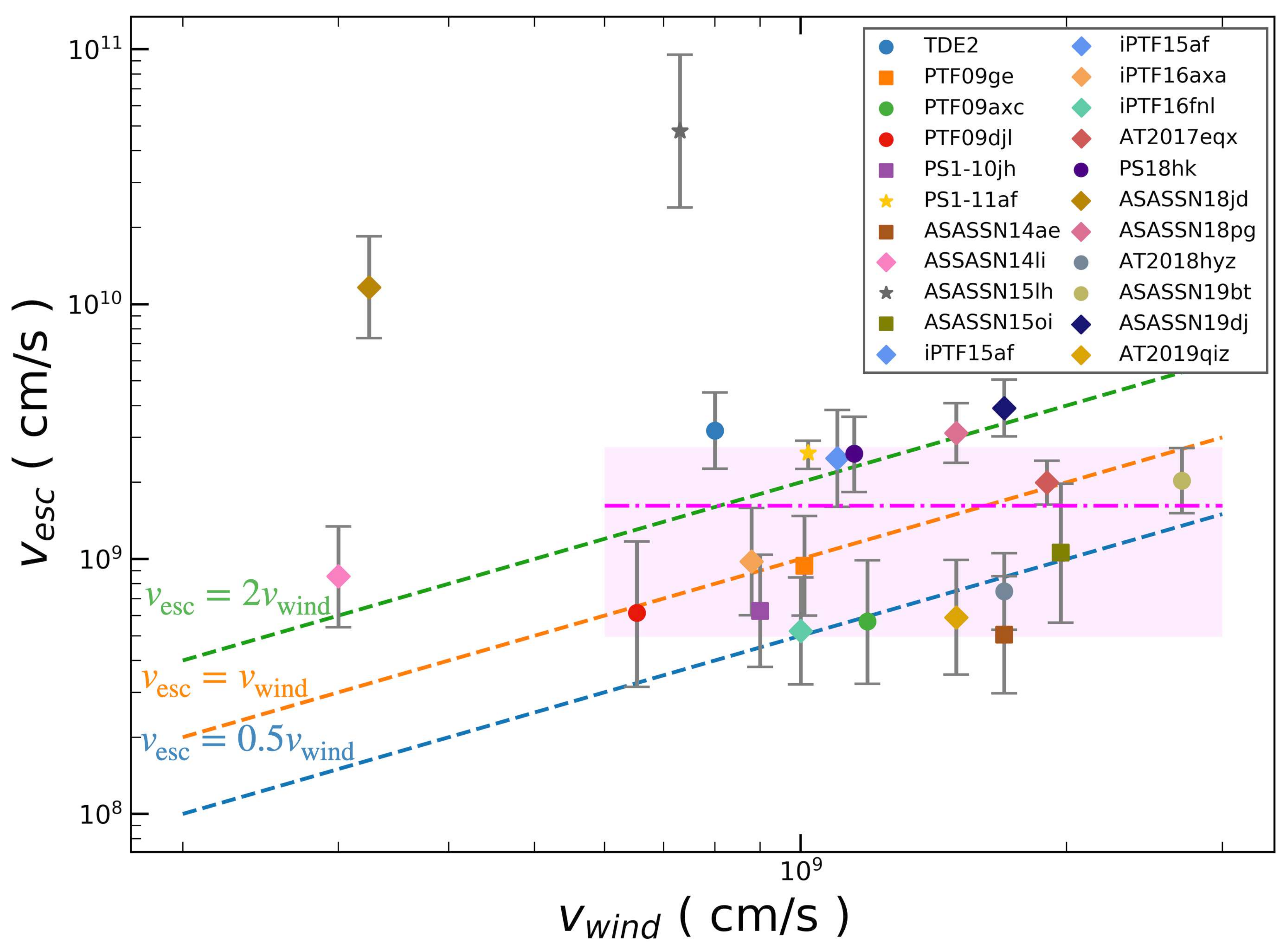}
\caption{Comparison between $v_{\rm wind}$ and $v_{\rm esc}$. The symbols are different for different spectral types of TDEs \citep{2020arXiv200101409V}. The filled circles, diamonds, squares, and stars show TDE-H, TDE-Bowen, TDE-He, and unspecified spectral type, respectively. The blue, orange, and green dashed lines show $v_{\rm esc} = 0.5 v_{\rm wind}$, $v_{\rm esc} = v_{\rm wind}$, and $v_{\rm esc} = 2 v_{\rm wind}$, respectively. The magenta dash-dot line shows the mean value of $v_{\rm esc}$ of the sample, excluding ASASSN-15lh, ASASSN-18jd, and ASASSN-14li (see the main text). The region enclosed by magenta shows the 1-sigma region ($\mu - \sigma \leq v_{\rm esc} \leq \mu + \sigma$).}
\label{fig:fig1}
\end{figure}

To test the assumption, $R_{\rm eq} = R_{\rm SI}$, we estimate $v_{\rm esc}$ at $R_{\rm SI}$, using $v_{\rm esc} = \sqrt{2GM_{\rm BH}/R_{\rm SI}}$. We regard the FWHM as $v_{\rm wind}$, and show the comparison between $v_{\rm esc}$ and $v_{\rm wind}$ in Figure\ref{fig:fig1}. 
Figure\ref{fig:fig1} shows that $v_{\rm esc}$ is roughly consistent with $v_{\rm wind}$ within a factor of 2. Discussing further the possible correlation is however difficult; how to accurately derive $v_{\rm wind}$ from observational data involves a large uncertainty. In addition, the present sample for this analysis is still limited. Indeed, the values of $v_{\rm wind}$ of most of the TDE samples here fall into a limited range within a factor of 3 (except for ASASSN-15lh, ASASSN-18jd, and ASASSN-14li, in which the former two are outliers). Therefore, the present sample would not allow such a detailed investigation of the correlation; the correlation coefficient between $v_{\rm esc}$ and $v_{\rm wind}$ is indeed smaller than 0.2, but this may simply be an outcome of the currently limited samples. Alternatively, we may simply discuss the mean and standard deviation in the distribution of $v_{\rm esc}$ (but excluding the above mentioned three objects). The mean and standard deviation are $\mu = 1.62 \times 10^{9}{\rm ~cm/s}$ and $\sigma = 1.12\times 10^{9}{\rm ~cm/s}$, respectively. This is roughly consistent with the above estimate within a factor of 2. In the future, we hope to analyze the possible correlation further, once the sufficiently large sample covering a range of $v_{\rm wind}$ becomes available and the relation between $v_{\rm wind}$ and the observed line width is better clarified.

In the two outliers; ASASSN-15lh and ASASSN-18jd, $v_{\rm wind}$ is significantly lower than $v_{\rm esc}$. It suggests that these objects are beyond the applicability of this model. Indeed, the observations show that ASASSN-15lh may be induced by an SMBH with mass above the upper limit to produce TDEs ($\sim 10^{8}~M_{\odot}$). ASASSN-18jd is not robustly identified as a TDE; we cannot dismiss the possibility that ASASSN-18jd is an active galactic nucleus or an unknown type of transients. We may need to consider different scenarios or emission mechanisms for these objects, including a possibility of a high BH spin \citep[][see also Section\ref{sec:4}]{2020MNRAS.497L..13M}.

In Figure\ref{fig:fig1}, different spectral types of TDEs \citep{2020arXiv200101409V} are shown by different symbols. No clear difference is seen in the distribution of $v_{\rm esc}$ and $v_{\rm wind}$ for different spectral types.

\subsection{Estimate of Physical Properties}\label{sec:3.2}

In \citet{2020ApJ...897..156U}, $R_{\rm eq}$ was one of the output parameters. However, in the present work, we treat $R_{\rm eq}$ as an input parameter under the assumption $R_{\rm eq} = R_{\rm SI}$. We alternately introduce a new parameter, $f$, into the equations.
The new parameter $f$ is the ratio of the kinetic energy ($\varepsilon_{\rm kin}$) to the thermal energy ($\varepsilon_{\rm th}$) per unit of volume at $R_{\rm SI}$. $f$ is described as $f  = \varepsilon_{\rm th}/\varepsilon_{\rm kin}$. 
In \citet{2020ApJ...897..156U}, it is assumed that $\varepsilon_{\rm th} = \varepsilon_{\rm kin}$ holds at $R_{\rm eq}$, but this time we incorporate the ratio as a new unknown parameter.
We expect that the derived value of $f$ should be an order of unity, if the model is self-consistent.

\citet{2020ApJ...897..156U} defines three typical physical scales; the wind-launched radius ($R_{\rm eq}$), the photon-trapped radius ($R_{\rm ad}$), and the color radius ($R_{\rm c}$). In this study, we replace $R_{\rm eq}$ by $R_{\rm SI}$. At $R_{\rm SI}$, we assume the following relation:
\begin{align}
  aT(R_{\rm SI})^4 = f \frac{1}{2}\rho(R_{\rm SI})v^2,
\end{align}
where the density structure, $\rho(r)$, is given as follows:
\begin{align}
  \rho(r) = \frac{\dot{M}}{4\pi r^2 v}.
\end{align}
We assume that $v_{\rm wind}$ is constant as a function of radius.
$R_{\rm ad}$ is defined by  $\tau_{\rm s}(R_{\rm ad}) = c/v$, where $\tau_{\rm s}$ is the optical depth for electron scattering. $R_{\rm c}$ is defined by $\tau_{\rm eff}(R_{\rm c}) = 1$, where $\tau_{\rm eff}$ is the effective optical depth, considering not only electron
scattering but also absorption processes.
The formation of the photosphere depends on a relative relation between $R_{\rm c}$ and $R_{\rm ad}$. The photospheric radius ($R_{\rm ph}$) is given as $R_{\rm ph} = \max (R_{\rm ad}, R_{\rm c})$. We also define the luminosity as follows:
\begin{align}
  L(r)=-\frac{4 \pi r^{2} a c}{3 \kappa_{\mathrm{s}} \rho} \frac{\partial}{\partial r} T^{4}.
\end{align}

Using above equations, we can estimate $f$ and $\dot{M}$. When $R_{\rm c} < R_{\rm ad}$, they are given as follows:
\begin{align}
\begin{split}
    \dot{M} &\approx 21 ~M_{\odot}{\rm yr^{-1}}\left(\frac{L}{1.0\times 10^{44}{\rm~ erg~s^{-1}}}\right)^{\frac{1}{2}}\\
    &\left(\frac{T_{\rm ph}}{3.0\times 10^{4}{\rm ~K}}\right)^{-2}\left(\frac{v}{9.0\times 10^{8}{\rm ~cm~s^{-1}}}\right)^{-\frac{1}{2}}, ~\rm and
\end{split}\\
\begin{split}
 f &\approx 0.29 \left(\frac{L}{1.0\times 10^{44}{\rm~ erg~s^{-1}}}\right)^{\frac{5}{6}}\left(\frac{T_{\rm ph}}{3.0\times 10^{4}{\rm ~K}}\right)^{\frac{2}{3}}\\
  &\left(\frac{v}{9.0\times 10^{8}{\rm ~cm~s^{-1}}}\right)^{-\frac{11}{6}}\left(\frac{R_{\rm SI}}{6.0\times 10^{14}{\rm ~cm}}\right)^{-\frac{2}{3}},
\end{split}
\end{align}
where $T_{\rm ph}$ is the photospheric temperature; $T_{\rm ph} = T(R_{\rm ph})$.
On the other hand, if $R_{\rm c} > R_{\rm ad}$ holds, the parameters are given as follows:
\begin{align}
\begin{split}
 \dot{M} &\approx 19 ~M_{\odot}{\rm /yr}\left(\frac{L}{8.0\times 10^{43}{\rm~ erg~s^{-1}}}\right)^{\frac{4}{5}}\\
 &\left(\frac{T_{\rm ph}}{2.5\times 10^{4}{\rm ~K}}\right)^{-\frac{11}{10}}
 \left(\frac{v}{6.5\times 10^{8}{\rm ~cm~s^{-1}}}\right), ~\rm and
\end{split}\\
\begin{split}
  f &\approx 0.53 \left(\frac{L}{8.0\times 10^{43}{\rm~ erg~s^{-1}}}\right)^{\frac{11}{15}}\left(\frac{T_{\rm ph}}{2.5\times 10^{4}{\rm ~K}}\right)^{\frac{11}{30}}\\
  &\left(\frac{v}{6.5\times 10^{8}{\rm ~cm~s^{-1}}}\right)^{-\frac{7}{3}}\left(\frac{R_{\rm SI}}{5.0\times 10^{14}{\rm ~cm}}\right)^{-\frac{2}{3}}.
\end{split}
\end{align}

\begin{figure}
\epsscale{1.17}
\plotone{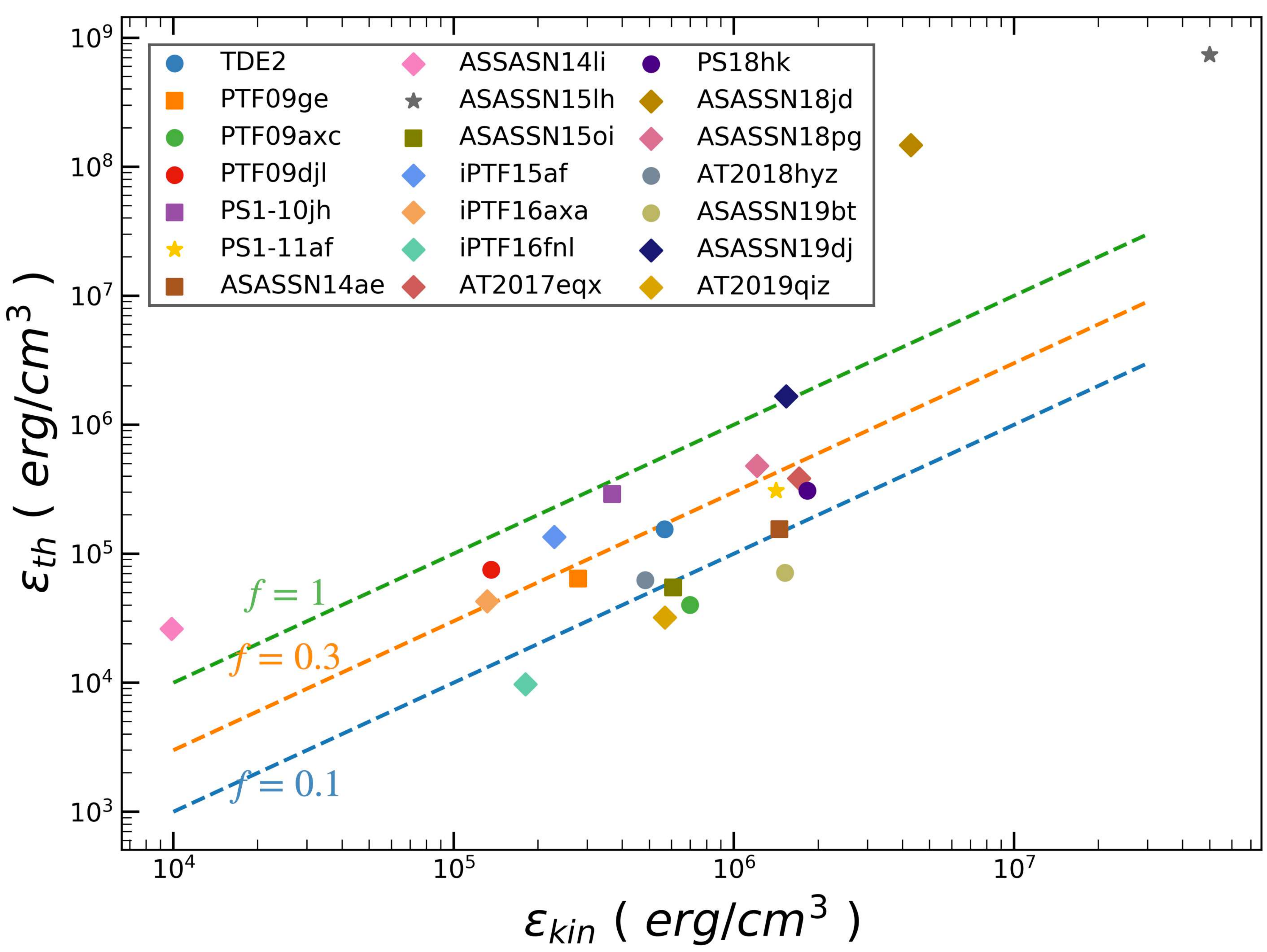}
\caption{Comparison between $\varepsilon_{\rm kin}$ and $\varepsilon_{\rm th}$. The blue, orange, and green dashed lines show $f = 0.1$, $f = 0.3$, and $f = 1$, respectively.}
\label{fig:fig2}
\end{figure}

Figure\ref{fig:fig2} shows the comparison between $\varepsilon_{\rm kin}$ and $\varepsilon_{\rm th}$ as we have derived. This shows a roughly positive correlation, but there are two objects which are clearly out of the trend; ASASSN-15lh and ASASSN-18jd. These are also the outliers in Figure\ref{fig:fig1}. This also suggests that they are beyond the applicability of the present model.
Generally, the outflow may well be launched from a radius where $\varepsilon_{\rm kin}$ and $\varepsilon_{\rm th}$ become comparable, i.e., $f \sim 1$ \citep{2009MNRAS.400.2070S}. 
Except for the two outliers, the ratio roughly stays in the range between $0.1$ and $1$. We suggest that this result is another strong support for $R_{\rm eq} = R_{\rm SI}$.

\begin{figure}
\epsscale{1.17}
\plotone{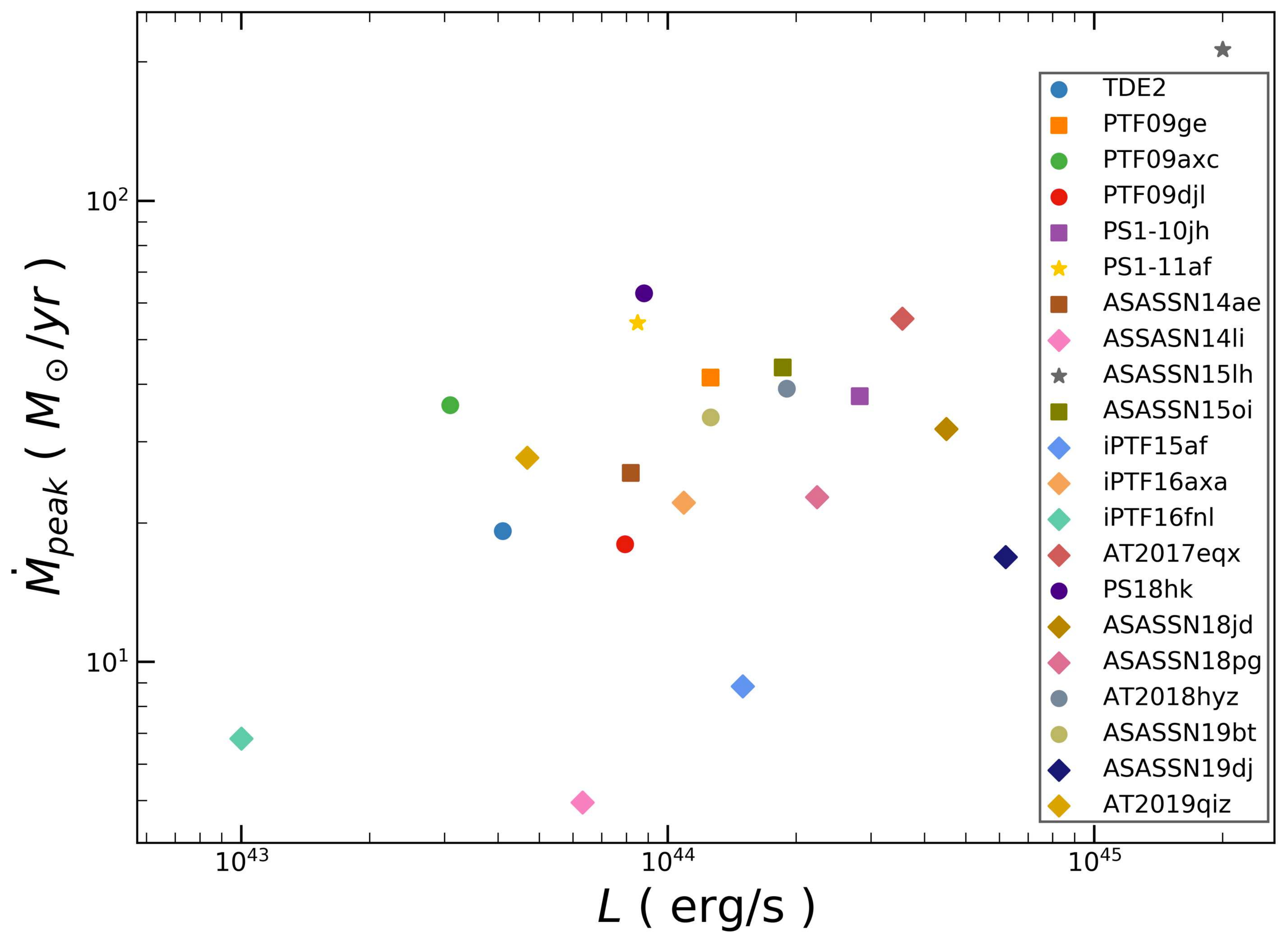}
\caption{$\dot{M}_{\rm peak}$ estimated by the Wind-Driven model.}
\label{fig:fig3}
\end{figure}

Figure\ref{fig:fig3} shows the estimated $\dot{M}_{\rm peak}$. We find that TDEs have strong outflows around the peak luminosity, typically with $\dot{M}_{\rm peak} \gtrsim 10~M_{\odot}{\rm /yr}$. Assuming that the disrupted star is $M_{*} \sim M_{\odot}$, they cannot release the mass exceeding $\sim 1M_{\odot}$. Extremely large mass-loss rates (e.g., ASASSN-15lh) are not feasible. This is another important constraint to identify the limits in the application of the model (see also Section\ref{sec:4}).

\begin{figure}
\epsscale{1.17}
\plotone{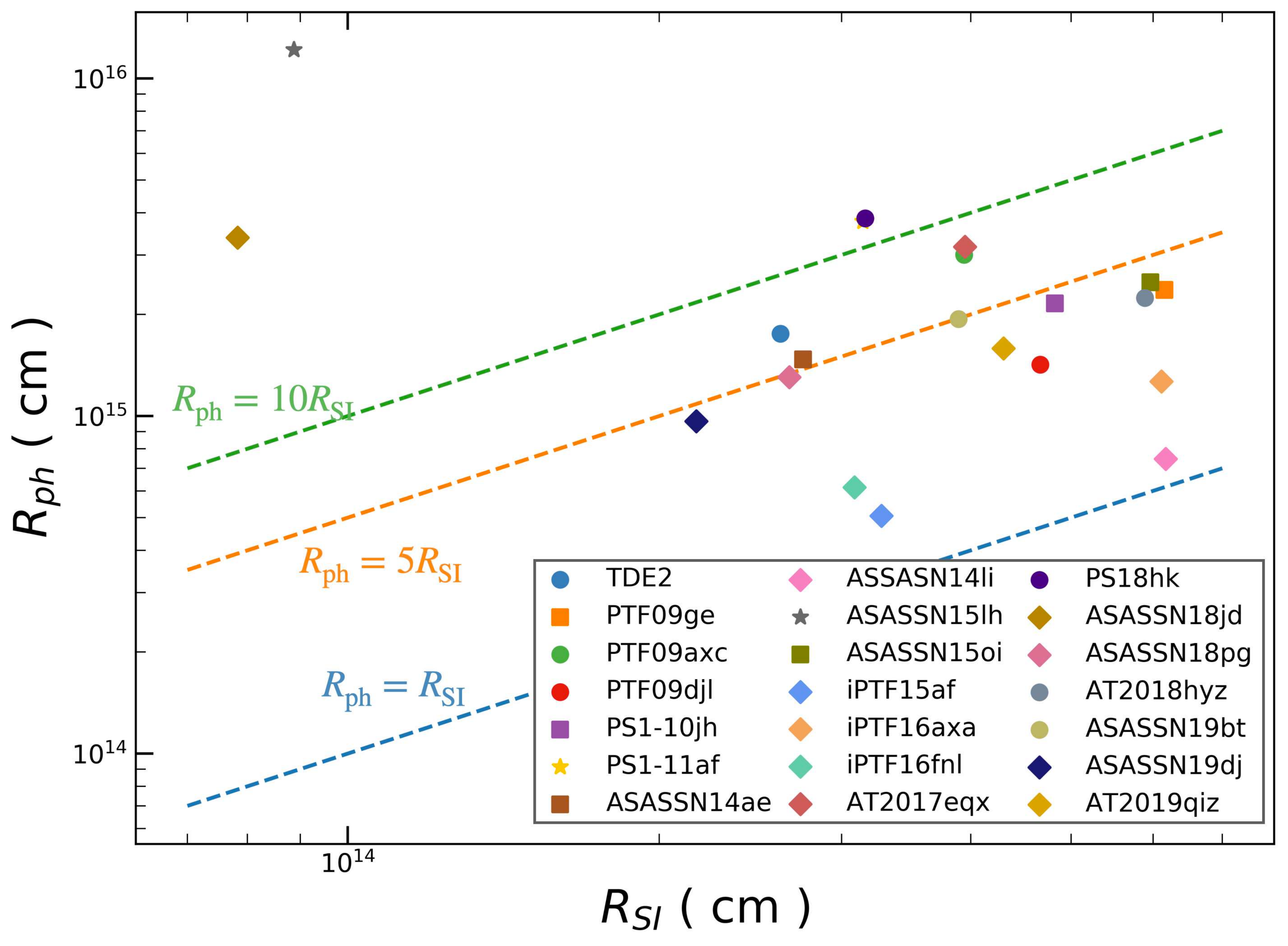}
\caption{Comparison between $R_{\rm SI}$ and $R_{\rm ph}$. The blue, orange, and green dashed lines show $R_{\rm ph} = R_{\rm SI}$, $R_{\rm ph} = 5R_{\rm SI}$, and $R_{\rm ph} = 10R_{\rm SI}$, respectively.}
\label{fig:fig4}
\end{figure}

Figure\ref{fig:fig4} shows the comparison between $R_{\rm SI}$ and $R_{\rm ph}$. Typically, $R_{\rm ph}$ is formed above $R_{\rm SI}$, between $\sim R_{\rm SI}$ and $\sim 10R_{\rm SI}$.
This result supports the picture that the origin of the Optical/UV radiation and some spectral lines is not the direct radiation from the accretion disk, but is the optically thick winds \citep{2009MNRAS.400.2070S}.

\section{Discussion}\label{sec:4}

We have generally derived high mass-loss rates for a sample of TDEs as an outcome of the wind-driven model for TDEs. Indeed,  TDEs are expected to have high accretion rates at peak. In \citet{2013MNRAS.435.1809S}, the peak fall-back rate, described as $\dot{M}_{\rm fb}$, is roughly given as follows:
\begin{align}
    \frac{\dot{M}_{\mathrm{fb}}}{\dot{M}_{\mathrm{Edd}}} \approx 133 \left(\frac{\eta}{0.1}\right) \left(\frac{M_{\rm BH}}{10^{6}M_{\odot}}\right)^{-\frac{3}{2}} \left(\frac{M_{*}}{M_{\odot}}\right)^{\frac{4}{5}},
\end{align}
where $\dot{M}_{\mathrm{Edd}}=4 \pi G M_{\rm BH} /\left(\kappa_{\mathrm{s}} \eta c\right)$, or,
\begin{align}
    \dot{M}_{\mathrm{fb}} \approx 3.45 ~ M_{\odot}{\rm /yr} \left(\frac{M_{\rm BH}}{10^{6}M_\odot}\right)^{-\frac{1}{2}} \left(\frac{M_{*}}{M_\odot}\right)^{\frac{4}{5}}.
\end{align}
Since this is the super-Eddington accretion, it is likely that most of the accretion may indeed be ejected (i.e., outflows) and therefore the accretion rate here may represent the mass-loss rate. Therefore, the high mass-loss rate we derived is in line with this expectation from the TDE physics. While it is true that the theoretically expected rate here is a few times smaller than our estimate, we note that both estimates involve uncertainties which would not allow detailed comparison.

Since TDEs are highly time-dependent transients, further discussion requires the effect of time evolution \citep[see, e.g., ][]{2020ApJ...897..156U}. One way to take this into account is to check the total mass ejected by the outflow. The timescale in the TDE light curve may differ from one object to another; some TDEs rapidly fade in $\sim 20$ days, while others stay bright for about a year \citep{2020arXiv200101409V}. With this caveat in mind, we adopt one month as the typical time scale, since most TDEs typically fade substantially in this timescale. By multiplying this time scale to the peak luminosities, the total ejected masses are expected to be a few $M_{\odot}$ (see also Figure\ref{fig:fig3}).

Since we assume $M_{*} = M_{\odot}$, the system cannot eject mass larger than $M_{\odot}$. However, $R_{\rm SI}$, which affects the mass-loss rates, depends on $M_{*}$ as follows: $R_{\rm SI} \propto R_{\rm t} \propto R_{*}M_{*}^{-1/3}$.
For larger disrupted stars, we thus expect that $R_{\rm SI}$ roughly stays the same because both $R_{*}$ and $M_{*}$ become larger. Therefore, considering more massive stars would not alter the present results substantially, and then the mass ejection of a few $M_{\odot}$ can be easily accommodated.

Indeed, \citet{2020arXiv200901240M} has recently estimated the mass ejection of some TDEs using the model similar to the present work. They have applied the model to a few TDEs, partially taking time evolution into account. They thereby derived $\sim 10M_{\odot}$ for the total ejected mass. This is roughly consistent with our estimate. With this value, they indeed have argued against the Wind-Driven model for TDEs. However, given that both models still lack detailed treatment of time evolution and also use some simplified assumption (e.g., treatment of opacity), and that we find a few other independent supports for the wind-driven model, we do take this rough agreement in the ejected mass between our estimate and the TDE expectation as another support for the applicability of the Wind-Driven model to TDEs.

As mentioned above, further detailed analysis will require the full time-dependent treatment. Before being able to sophisticate the model at this level, we indeed need to overcome several limitations in our current understanding of the nature of TDEs in their observational data. For example, our current understanding of the evolution of $v_{\rm wind}$ is not sufficient, which requires deeper understanding of the line formation processes \citep[see, e.g., ][]{2020ApJ...897..156U}. Also, a sample of TDEs with well-sampled time evolution data is still limited. 
Therefore, in this work, we focus on the properties of TDEs at their peaks, aiming at understanding the general/statistical properties of TDEs using as large an observed sample as is currently available.
In the future, we hope to address the time evolution effects and apply such a model to a sample of TDEs with the time-evolution data available, but this is beyond a scope of the present work.

\citet{2020MNRAS.497L..13M} presented that ASASSN-15lh is a peculiar TDE which has a high BH spin close to $a_{\rm BH} \approx 0.99$. Adopting the high BH spin, we have applied our model to ASASSN-15lh. The main change is seen in the value of $f$, but this is only about 2\%. Taking into account the BH spin thus does not have a significant effect on the derived properties based on the wind model, e.g., mass-loss rate. ASASSN-15lh thus remains an outlier, which seems to be beyond the applicability of our present model.

Using the Wind-Driven model and the results, we can obtain insights for TDEs or other astronomical transients driven by explosive mass ejection. Namely, we can constrain some physical properties for transients with insufficient observational data. For example, we can roughly estimate $M_{\rm BH}$ or $v_{\rm wind}$.

As one such application, we estimate $M_{\rm BH}$ for a peculiar transient, AT2018cow \citep{2018ApJ...865L...3P}. 
AT2018cow is a luminous blue transient ($L_{\rm peak} \approx 4\times 10^{44}{\rm ~erg~s^{-1}}$, $T_{\rm peak} \approx 31400{\rm ~K}$, and $v \approx 0.1c$) discovered by ATLAS on MJD 58285 \citep{2019MNRAS.484.1031P}. In \citet{2019MNRAS.484.1031P}, they argued that AT2018cow is a TDE induced by an intermediate-mass black hole (IMBH). However, AT2018cow occurred far from the galactic center, which makes it difficult to estimate $M_{\rm BH}$ from the $M_{\rm bulge}-M_{\rm BH}$ relation. They estimated $M_{\rm BH}$ using Mosfit TDE model \citep{2018ApJS..236....6G}. In our model, we can constrain $M_{\rm BH}$ independently from Mosfit. Assuming $R_{*} = R_\odot$, $M_{*} = M_{\odot}$, $\beta = 1$, and $f = 0.5$, we estimate $M_{\rm BH}$ as $\sim 2.5 \times 10^{4} ~M_{\odot}$. This is consistent with the estimate by \citet{2019MNRAS.484.1031P}. Our model thus supports that AT2018cow may be a TDE induced by an IMBH. In addition, we note that $R_{\rm SI}$ estimated by \citet{2019MNRAS.484.1031P} is a factor of 10 smaller than the observed photosphere. This is also consistent with our result (see Figure\ref{fig:fig4}), supporting by the Wind-Driven model.

\section{Conclusions}\label{sec:5}

Using the Wind-Driven model presented by \citet{2020ApJ...897..156U}, we have aimed to constrain the origin of the optical/UV radiation in TDEs. The comparison between the escape velocities and the wind velocities supports that the wind is launched from the self-interaction radius.
Generally, the wind is expected to be launched from a position where the ratio of kinetic to thermal energy per unit volume is roughly equal (i.e., equipartition). We also estimate the ratio at the self-interaction radius through the Wind-Driven model, and it turns out to be an order of unity. This result supports the assumption that the stream collision induces the wind.

We find TDEs have strong outflow around the peak. The mass-loss rates are typically over $10 ~M_{\odot}{\rm yr^{-1}}$. We also show that the photospheric radii are 1-10 times larger than the self-interaction radii. This result supports the picture that the Optical/UV radiation is emitted not from the accretion disk directly, but from an optically thick wind.

Applying the framework to TDEs or other astronomical transients driven by explosive mass ejections, we can obtain constraints on the physical properties that can not be obtained from observations. 
We apply the framework to a peculiar transient, AT2018cow. The model suggests that AT2018cow is likely a TDE induced by an intermediate-mass black hole ($\sim 10^{4}~M_{\odot}$).

The Wind-Driven model still has significant room for improvement. The present model assumes a steady-state, and in this Letter we estimate physical quantities only at the peak. To obtain detailed constraints, it is necessary to create a non-steady-state model that takes into account the time evolution and radial dependence of the velocity. We postpone such work to the future.

\acknowledgments
We thank Girogs Leloudas and Tatsuya Matsumoto for helpful comments and discussions.
K.M. acknowledges support provided by Japan Society for the Promotion of Science (JSPS) through KAKENHI grant (18H05223, 20H00174, and 20H04737).

\bibliography{manuscript}{}
\bibliographystyle{aasjournal}

\end{document}